\documentclass[10pt,aps,prb,twocolumn,amsmath,amssymb,superscriptaddress]{revtex4-1}
\usepackage{amsmath}
\usepackage{graphicx}
\usepackage{color}
\usepackage{multirow}
\usepackage[caption = false]{subfig}

\newcommand{\ket}[1]{\left|  #1  \right\rangle}
\renewcommand\b[1]{\ensuremath{\mathbf{#1}}}
\newcommand\B[1]  {\ensuremath{\pmb #1}}

\usepackage{array}
\newcolumntype{P}[1]{>{\centering\arraybackslash}p{#1}}
\newcolumntype{L}[1]{>{\raggedright\arraybackslash}p{#1}}
\newcolumntype{R}[1]{>{\raggedleft\arraybackslash}p{#1}}

\newcommand\equ[2]{\begin{align}#2\label{#1}\end{align}}

\begin{document}
\renewcommand{\arraystretch}{1.2}

\title{One-step treatment of spin-orbit coupling and electron correlation in large active spaces}
%: calculation of g-tensor in heavy transition metal atoms}
%\title{Treating large active spaces cheaply using a stochastic Heat-bath Configuration Interaction}
%\title{Efficient stochastic perturbation theory in Heat-bath Configuration Interaction}
%\title{Efficient treatment of large active spaces via stochastic perturbation theory in Heat-bath Configuration Interaction}
%\title{Corrections to selected configuration interaction energies using stochastic multireference Epstein-Nesbet perturbation theory}
\author{Bastien Mussard}
\email{bastien.mussard@colorado.edu}
\affiliation{Department of Chemistry and Biochemistry, University of Colorado Boulder, Boulder, CO 80302, USA}
\author{Sandeep Sharma}
\email{sanshar@gmail.com}
\affiliation{Department of Chemistry and Biochemistry, University of Colorado Boulder, Boulder, CO 80302, USA}
\begin{abstract}
In this work we demonstrate that the heat bath configuration interaction (HCI) and its semistochastic extension can be used to treat relativistic effects and electron correlation on an equal footing in large active spaces to calculate the low energy spectrum of several systems including halogens group atoms (F, Cl, Br, I), coinage atoms (Cu, Au) and the Neptunyl(VI) dioxide radical. This work demonstrates that despite a significant increase in the size of the Hilbert space due to spin
symmetry breaking by the spin-orbit coupling terms, HCI retains the ability to discard large parts of the low importance Hilbert space to deliver converged absolute and relative energies. For instance, by using just over $10^7$ determinants we get converged excitation energies for Au atom in an active space containing (150o,25e) which has over $10^{30}$ determinants. We also investigate the accuracy of five different two-component relativistic Hamiltonians in which different levels of approximations are made in deriving the one-electron and two-electrons Hamiltonians, ranging from Breit-Pauli (BP) to various flavors of exact two-component (X2C) theory. The relative accuracy of the different Hamiltonians are compared on systems that range in atomic number from first row atoms to actinides. 
\end{abstract}
\maketitle

\section{Introduction}
Relativistic effects play an important role in a variety of photochemical processes, various magnetic spectroscopies and are responsible for unusual phase behaviors in transition metal oxides. In practice, the most accurate procedure for treating relativistic effects in molecular systems is through the solution of the Dirac-Coulomb-Breit (DCB) equation\cite{Bar-Chapter,Liu2010,Fleig2012,Saue2011,Kut-CP-2012,Pyy-ARPC-2012,Pyy-CR-2012}. Although recent developments have made it possible to perform self-consistent field\cite{Pernpointner2000,Yanai2001,Thyssen2008,Kelley2013}, density functional theory\cite{Yanai2001a,Saue2002,Belpassi2011,Mizukami2011}, coupled cluster\cite{Nataraj2010,Sorensen2011}, explicitly correlated\cite{Bylicki2008,Li2011,Ten-no2012} and multireference\cite{Almoukhalalati2016,Bates2015,Fleig2003,Fleig2006,Knecht2010,Shiozaki2015} calculations on the DCB Hamiltonian, these methods remain problematic due to their high computational cost.

An alternative is to use two-component Hamiltonians which are capable of delivering quantitative accuracy for relativistic problems. They are derived by either eliminating the small component (ESC) or by using a unitary transformation called the Foldy-Wouthuysen (FW) transformation\cite{FolWou-PR-1950} to decouple the large and the small components in the one-body Dirac or Dirac-Fock Hamiltonian. These two are of course related and FW transformation can be viewed as a normalized elimination of small component (NESC). This transformation is also applied to the two-electron integrals to generate a relativistically correct electron interaction. The resulting two-component Hamiltonian can be partitioned into the Schr\"{o}dinger equation and additional relativistic spin-free and spin-dependent terms. The Schr\"{o}dinger equation together with the spin-free terms are referred to as the scalar relativistic Hamiltonian and calculating the ground state energy of this Hamiltonian is usually no more expensive than that of
the Schr\"{o}dinger equation. The remaining spin-dependent terms, known as the spin-orbit coupling (SOC), are usually treated using perturbation theory. However, for heavier elements these terms can become large and it is more appropriate to treat them on an equal footing with the spin-free terms and electron correlation. In this work we follow the latter approach.

The three most commonly used two-component Hamiltonians are the Douglass-Kroll-Hess\cite{Douglas1974,PhysRevA.33.3742,Nakajima2012,Reiher2006}, the Barysz-Sadlej-Snijders\cite{Barysz2002,Barysz2009,Ilias2005} and the exact two-component (X2C) Hamiltonians\cite{Dyall1997,Kutzelnigg2005,Liu2007,Ilias2007}. Both Douglass-Kroll-Hess and Barysz-Sadlej-Snijders Hamiltonians carry out the transformation analytically. The transformation operator contains a complicated function of the momentum operator and its integrals cannot be calculated analytically, instead the matrix representation of the momentum operator in a finite basis is calculated. In X2C, one forgoes the analytic transformation and the entire process is carried out algebraically using the matrix representation of operators, starting from the solution of the Dirac equation. The commonly used approximation, known as X2C-1e, consists of solving the spin-free non-interacting Dirac equation in one step (without electron-electron interaction a self-consistent procedure is not needed) and the transformation matrix is derived from this solution. The motivation for doing so comes from the fact that only a small fraction of the total cost of performing a correlated calculation is used in the solution of the one-body problem and thus solving it to exactly decouple the large and small component is relatively cheap.
	
To get quantitative accuracy one also needs to include the relativistically correct two-body terms. Such terms are derived from the transformation of the Coulomb and Gaunt operators, which gives rise to several spin-free and spin-dependent terms. In this work we will ignore the relativistic correction to the spin-free two-body terms and only some of the spin-dependent terms are included that appear at order $\alpha^{2}$. The effect of those two-body SOC terms is often treated approximately using the spin-orbit mean field
(SOMF) approximation\cite{Hessetal,Marian1996}, which replaces the two-body terms by an effective fock-like one-body term. Although the approximation sounds drastic, in practice it is known to be extremely accurate and is almost universally used. The SOMF approximation simplifies the correlated calculation considerably by reducing the memory and CPU requirement of storing and transforming the different sets of relativistic two-electron integrals.

Several algorithms and programs are now available for performing relativistic calculations with two-component Hamiltonians\cite{Buenker1998,BERNING2000,Vallet2000,Yanai2003,Gordon2005,Kleinschmidt2006,Gauss2009,Neese2012,Mai2014,MOLPRO_brief,Aquilante2016}. Some of these methods include the spin-orbit coupling in the self-consistent field (SCF) calculation, due to which the orbital relaxation effects are fully included, but the resulting orbitals are complex-valued spinors\cite{Esser1981,MyeongCheolKim1996,Fleig2001,Kim2014,Kim2012}. In the intermediate approach the spin-orbit coupling terms are only introduced during the correlated calculation\cite{Sjovoll1997,Yabushita1999,Fedorov2000,Kleinschmidt2006,Ganyushin2013}, and are treated on an equal footing with the dynamical electron correlation. A further approximation is possible which is usually called the quasi-degenerate perturbation theory or the state interaction approach, whereby one first calculates several electronic state wavefunctions of the spin-free Hamiltonian in different spin sectors.
The matrix representation in the basis of these states of the complete two-component Hamiltonian including spin-orbit coupling is then diagonalized to obtain the spin-orbit coupled results\cite{B401472N,Ganyushin2006,Roemelt2015,Sayfutyarova2016,Knecht2016}.
The accuracy of the method is limited by the number of states included in the state interaction approach, which is usually on the order of a few tens of states and in rare cases of heavy elements a few hundred\cite{Liu2016}. A related approach is the EOM(SOC)-CC\cite{Klein2008,Cao2017} where the spin-orbit coupling terms are included only during the equation of motion part of the calculation and thus are effectively treated perturbatively.

In this work we perform multireference calculations in large active spaces where the SOC is treated on an equal footing as the electron correlation. Heat-bath configuration interaction (HCI) algorithm\cite{HolTubUmr-JCTC-16,ShaHolUmr-JCTC-17} is the multireference method which is used to calculate the zero-field splittings using real-valued orbitals obtained from a non-relativistic SCF calculation. HCI is an efficient variant of the general class of methods in which a selected configuration
interaction is performed which is followed by perturbation theory. As we will show in the results section, HCI is able to systematically discard large parts of the Hilbert space to deliver accurate excitation energies at a much reduced cost relative to a full configuration interaction (FCI) calculation. To the best of our knowledge current calculations and the density matrix renormalization group (DMRG) theory calculations performed by Sayfutyarova \textit{et al.}\cite{Sayfutyarova2016} and
Knecht \textit{et al.}\cite{Knecht2014} are the only ones in literature where relativistic calculations are performed on an active space larger than 16 electrons in 16 orbitals. However, unlike the DMRG calculations of Sayfutyarova, here we treat SOC non-perturbatively on an equal footing with the electron correlation.

The rest of the article is organized as follows.
In Section II, we present the derivation and the working equations of the Breit-Pauli and X2C Hamiltonians used in this work. In Section III, we describe the Heat-bath Configuration Interaction algorithm and the extensions to it that allow us to treat the SOC terms.
In Section IV and V, we give the computational details and results respectively.% on the workflow used for the calculations of the zero-field splitting done in this work is explained, and Section V presents results of such calculations on the atoms of the halogen group, on the copper and gold atoms and on the Neptunium dioxide.

\section{Theory}

In this section, we present the working equations of the two-component Hamiltonians used in this work: the Breit-Pauli (BP) and the exact two-component (X2C) Hamiltonians.
To give some context, we start with the matrix formulation of the one-electron four-component Dirac Hamiltonian
\equ{eq:Dirac}{
   \begin{pmatrix}
       \b{V}_\text{ne} & c\;\B{\sigma}\cdot\b{p} \\
       c\;\B{\sigma}\cdot\b{p} & \b{V}_\text{ne}- 2 m c^2
   \end{pmatrix}
   \Psi
   =E\Psi
,}
where $\b{V}_\text{ne}$ is the  electron-nucleus potential,
$\b{p}$ is the momentum and $\B{\sigma}=\{\B{\sigma}^x,\B{\sigma}^y,\B{\sigma}^z\}$ is the set of Pauli matrices.
The eigenvectors $\Psi$ of the Dirac Hamiltonian are four-component bispinors which contain the
``large'' $\Psi^\text{L}$ and ``small'' $\Psi^\text{S}$ components that are themselves two-component wavefunctions:
\equ{ }{
    \Psi=\begin{pmatrix}\Psi^\text{L}\\\Psi^\text{S}\end{pmatrix}
.}
Although the Dirac Hamiltonian is fully Lorentz invariant, introducing the Coulomb interaction between the electrons breaks this invariance and one would need to go to quantum electrodynamics (QED) to obtain a fully Lorentz invariant theory of interacting electrons, but it is not obvious that a fully Lorentz invariant many-electron Hamiltonian can be derived from QED. In practice, the Dirac-Coulomb-Breit Hamiltonian with the  non-retarded electron-electron interaction
\begin{align}
\hat{V}_\text{ee}(i,j) &= \left(\frac{1}{r_{ij}}\right) - \left(\frac{\B{\alpha}_i\cdot\B{\alpha}_j}{r_{ij}}\right) 
\nonumber\\&\qquad\quad+ \left(\frac{\B{\alpha}_i\cdot\B{\alpha}_j}{2r_{ij}} - \frac{(\B{\alpha}_i\cdot\b{r}_{ij})(\B{\alpha}_j\cdot\b{r}_{ij})}{2r_{ij}^3} \right)\label{eq:cgb}
\end{align}
is used (in brackets are the Coulomb, Gaunt and Gauge terms, while $\B{\alpha}_i$ are the Dirac matrices) and is expected to be sufficiently accurate for chemical applications. 

In principle, it is possible to obtain exact two-component Hamiltonians by solving the DCB Hamiltonian and using the many-body solutions to decouple the small and the large components exactly. This is of course impractical because the full solution to the DCB Hamiltonian would be required. Instead, one tries to decouple the large and small components at the one-body level, which can be done approximately analytically (as is done to derive the Breit-Pauli Hamiltonian) or exactly algebraically (as is done to derive the exact two-component Hamiltonian). This class of procedures is tractable because one needs to solve at most the one-body problem. The transformation obtained this way is then applied to the two-body Coulomb, Gaunt and Breit terms to derive two-component two-body SOC Hamiltonians.
 
\subsection{Breit-Pauli Hamiltonian}
The Breit-Pauli (BP) Hamiltonian used in this work is a two-component Hamiltonian 
obtained
through an analytic FW transformation performed
on the four-component one-body Dirac Hamiltonian of Eq.~(\ref{eq:Dirac})
with two-body terms obtained by contributions from the transformed Coulomb and Gaunt terms (first and second terms of Eq.~(\ref{eq:cgb})).
The final expression of the Breit-Pauli Hamiltonian\cite{dyallbook,reiherbook} is
\equ{ }{
    \b{H}^\text{BP}=\b{H}^\text{BP}_\text{SF}+ \b{H}^\text{BP}_\text{SO}{}^{(1)} + \b{H}^\text{BP}_\text{MF}{}^{(1)}
,}
where the three terms are defined in Eqs.~(\ref{eq:BPsf}), (\ref{eq:BPso1}) and (\ref{eq:BPsomf}).

\subsubsection{One-body part}
The one-body part of the BP Hamiltonian, known as the Pauli Hamiltonian, contains the kinetic energy, electron-nuclear interaction, one-body spin-orbit coupling, mass-velocity and Darwin terms. In this work we will ignore the divergent mass-velocity and Darwin terms and are thus left with $\b{H}^\text{Pauli}=\b{H}^\text{BP}_\text{SF}+\b{H}^\text{BP}_\text{SO}{}^{(1)}$, where the spin-free part
\begin{align}
\b{H}^\text{BP}_\text{SF}=\b{T} + \b{V}_\text{ne}
\label{eq:BPsf}
\end{align}
contains the kinetic energy
$\b{T}$ and the electron nuclear attraction $\b{V}_\text{ne}$. The one-body spin-orbit coupling term is
\begin{align}
    \b{H}^\text{BP}_\text{SO}{}^{(1)}=-i\frac{\alpha^2}{4} \sum_{iA} \left(\b{p}_i\frac{Z_{A}}{r_{iA}}\times \b{p}_i\right)\cdot  \B{\sigma}_i
\label{eq:BPso1}
,\end{align}
where,
$\alpha$ is the fine structure constant,
$\b{p}_i$ is the momentum of electron $i$,
$Z_{a}$ is the atomic number of nucleus $A$,
and $r_{iA}$ the distance between electron $i$ and nucleus $A$.
In second quantization the one-body SOC term can be written as
\begin{align}
    \b{H}^\text{BP}_\text{SO}{}^{(1)}=&-i\frac{\alpha^2}{4}\sum_{\substack{iA\\pq\\\mu\nu\lambda}} \epsilon_{\mu\nu}^\lambda \langle p_\mu|\frac{Z_{A}}{r_{iA}} | q_\nu\rangle\hat{\sigma}_{pq}^\lambda\\
    =&-i\frac{\alpha^2}{4}\sum_{pq\lambda} {}^\text{BP}\!S_{pq}^\lambda\hat{\sigma}_{pq}^\lambda
    \label{eq:Sbp}
,\end{align}
where
$\lambda,\mu,\nu$ can be $(x,y,z)$,
$p$ and $q$ are orbitals,
$p_\mu  = \frac{\partial p}{\partial \mu}$ are nuclear derivatives of the orbitals,
and $\epsilon_{\mu\nu}^\lambda$ is the Levi-Civita symbol.
Note that the three $\hat{\sigma}$ operators are
\begin{align}
    \hat{\sigma}_{pq}^x &=   \hat{a}_{p\alpha}^\dag \hat{a}_{q\beta}  + \hat{a}_{p\beta}^\dag \hat{a}_{q\alpha} \nonumber\\
    \hat{\sigma}_{pq}^y &=-i(\hat{a}_{p\alpha}^\dag \hat{a}_{q\beta}  - \hat{a}_{p\beta}^\dag \hat{a}_{q\alpha})\\
    \hat{\sigma}_{pq}^z &=   \hat{a}_{p\alpha}^\dag \hat{a}_{q\alpha} - \hat{a}_{p\beta}^\dag \hat{a}_{q\beta}\nonumber
.\end{align}
We can infer from the definition of ${}^\text{BP}\!S_{pq}^\lambda$ that the one-body spin-orbit coupling integrals are anti-hermitian.

\subsubsection{Two-body part}
The application of the FW transformation to the Coulomb and Gaunt operators spawns several spin-free and spin-dependent terms, out of which in this work we will only include the spin-same-orbit and spin-other-orbit interactions, \textit{i.e.} respectively the first and second term in
\begin{align}
    \b{H}^\text{BP}_\text{SO}{}^{(2)}
    =& i \frac{\alpha^2}{4}\sum_{i\neq j} \left(\b{p}_i\frac{1}{r_{ij}}  \times\b{p}_i\right) \cdot (\B{\sigma}_i+2\B{\sigma}_j).
\end{align}
In second quantization, this takes the form
\begin{align}
    \b{H}^\text{BP}_\text{SO}{}^{(2)}
    =&i \frac{\alpha^2}{4}\sum_{pqrs\lambda}\left( 2J_{rspq}^\lambda +  J_{pqrs}^\lambda\right) \hat{D}^\lambda_{pqrs}
,\end{align}
where the two-electron integrals are
\begin{align}
    J_{pqrs}^\lambda=\sum_{\mu\nu} \epsilon_{\mu\nu}^\lambda \langle p_\mu r| q_\nu s\rangle
    \label{eq:Jbp}
,\end{align}
and the operator is 
\begin{align}
\hat{D}^\lambda_{pqrs}=\hat{\sigma}_{pq}^\lambda \hat{E}_{rs}-\delta_{qr} \hat{\sigma}_{ps}^\lambda
.\end{align}
For real orbitals one can utilize the 4-fold symmetry 
\begin{align}
J_{pqrs}^\lambda = -J_{qprs}^\lambda=-J_{qpsr}^\lambda=J_{pqsr}^\lambda
\end{align}
to reduce the memory cost of storing the two electron integrals.
%Thus if the nuclear gradient derivatives are available, the spin-orbit coupling integrals can be calculated by performing appropriate summations.

Note that the two-body spin-orbit coupling term is of opposite sign to the one-body term of Eq.~(\ref{eq:Sbp}) and acts as a screening potential.
Several other terms including the spin-spin dipole interactions and Fermi contact interaction are ignored. For most applications such simplifications are justified, however some noteworthy exceptions exist such as the oxygen molecule\cite{Langhoff1974,Vahtras2002}, where the spin-spin dipole interactions are of the same order of magnitude as the spin-orbit coupling terms.

Although the two-body terms can be easily incorporated in a calculation, the cost of storing and transforming integrals for large molecules become too expensive. This motivates the use of the spin-orbit mean field approximation\cite{Hessetal,Marian1996,Li2014}, where one writes an effective one-body operator by integrating out the spins of two out of the four orbitals in the two-electron integrals of Eq.~(\ref{eq:Jbp}).
Such an integration leads to the effective one-body term\cite{Hessetal} written in second quantization as
\begin{align}
    \b{H}^\text{BP}_\text{MF}{}^{(1)}=i\frac{\alpha^2}{4} \sum_{pq} {}^\text{BP}\!\bar{S}^\lambda_{pq} \hat{\sigma}_{ps}^\lambda,
\label{eq:BPsomf}
\end{align}
where
\begin{align}
    ^\text{BP}\!\bar{S}^\lambda_{pq}
    &=  \sum_{rs} \gamma_{rs} \left(J_{pqrs}^\lambda -\frac{3}{2} J_{prsq}^\lambda-\frac{3}{2} J_{sqpr}^\lambda\right)
     \label{eq:Sbpmf}
\end{align}
and $\gamma$ is the one-body density matrix.

Although it is not used here, the cost of calculating the two-body terms can be further substantially reduced by using the one-center approximation which makes use of the local nature of the spin-orbit interaction and introduces only a small error\cite{amfi}.

\subsection{X2C Hamiltonian}
The derivation of the X2C Hamiltonian begins by using the restricted kinetic balance\cite{StaHav-JCP-1984} that replaces the small component ($\psi^\text{S}$) with the pseudo-large component ($\phi^\text{L}$)
\begin{align}
\psi^\text{S}=\frac{\B{\sigma}\cdot\b{p}}{2c}\phi^\text{L}
,\end{align}
which gives rise to the modified four-component one-electron Dirac equation
\begin{align}
    \begin{pmatrix}
        \b{V}_\text{ne}& \b{T}\\
\b{T}& \b{W} - \b{T}\\
\end{pmatrix} \begin{pmatrix}\psi^\text{L}\\ \phi^\text{L}\end{pmatrix} = \begin{pmatrix}
    \b{S}& \b{0}\\
    \b{0}& \frac{\alpha^2}{4}\b{T}\\
\end{pmatrix} \begin{pmatrix}\psi^\text{L}\\ \phi^\text{L}\end{pmatrix}E
\label{eq:modDirac}
,\end{align}
where $\b{W} = \frac{\alpha^2}{4}(\B{\sigma}\cdot\b{p}) V_\text{ne} (\B{\sigma}\cdot\b{p})$ and $\b{S}$ is the non-relativistic metric. The restricted kinetic balance transformation ensures that the large and the pseudo-large components have the same symmetry properties and can be expanded with a common basis set. Also, the two components become equal to each other in the non-relativistic limit.

The only spin-dependent term in the modified Dirac equation of Eq.~(\ref{eq:modDirac}) is found in $\b{W}$, which can be exactly partitioned into a spin-free and spin-dependent term
\begin{align}
    \b{W}
    =& \frac{\alpha^2}{4}\b{p}\cdot V_\text{ne}\b{p} + i\frac{\alpha^2}{4}\B{\sigma}\cdot\left(\b{p} V_\text{ne}\times\b{p}\right)\nonumber\\
   =& \b{W}_\text{SF} + \b{W}_\text{SD}\label{eq:sd}
.\end{align}
This allows us to separate the spin-free ($\b{H}^\text{4c}_\text{SF}$) and spin-dependent ($\b{H}^\text{4c}_\text{SD}$) terms in the four-component modified Dirac equation 
\begin{align}
&\b{H}^\text{4c}_\text{SF} = \begin{pmatrix}
    \b{V}_\text{ne}& \b{T}\\
    \b{T}& \b{W}_\text{SF} - \b{T}
\end{pmatrix}\label{eq:sf}\\
&\b{H}^\text{4c}_\text{SD} = \begin{pmatrix}
    \b{0}& \b{0}\\
    \b{0}& \b{W}_\text{SD}
\end{pmatrix}.
\end{align}

An FW transformation can be derived by solving only the spin-free part of the one-body Hamiltonian (Eq.~\ref{eq:sf}) to obtain the X2C-1 one-body SOC terms\cite{Li2012,Li2013a,Li2014}. Another possibility is to obtain a complex-valued FW transformation by solving the total one-body Hamiltonian (containing the total $\b{W}$ of Eq.~\ref{eq:sd}) to obtain the X2C-N one-body SOC terms\cite{Cheng2014}. In both cases the two-body part is obtained by applying the X2C-1 FW transformation to the Coulomb and Gaunt operators.
The final expression of the resulting exact two-component Hamiltonians used in this work is thus:
\begin{align}
    \b{H}^\text{X2C-x}=\b{H}^\text{X2C}_\text{SF}+ \b{H}^\text{X2C-x}_\text{SO}{}^{(1)} + \b{H}^\text{X2C}_\text{MF}{}^{(1)}\label{eq:x2ceq}
,\end{align}
where ``x'' distingishes between the X2C-1 and X2C-N one-body terms. The $\b{H}^\text{X2C}_\text{SF}$ is defined in Eq.~(\ref{eq:X2Csf}), $\b{H}^\text{X2C-1}_\text{SO}{}^{(1)}$ is defined in Eq.~(\ref{eq:X2Cso1}), $\b{H}^\text{X2C-N}_\text{SO}{}^{(1)}$ is defined in Eq.~(\ref{eq:X2Cson}) and the $\b{H}^\text{X2C}_\text{MF}{}^{(1)}$ is defined in Eq.~(\ref{eq:X2Csomf}).

\subsubsection{One-body part: X2C-1 scheme}

As stated above, in the X2C-1 scheme one uses only the spin-free four-component Hamiltonian $\b{H}^\text{4c}_\text{SF}$ to solve the modified Dirac equation of Eq.~(\ref{eq:modDirac}) in a single step.
The data of the obtained four-component positive energy solutions $\Psi_+$
is used to apply an FW transformation to block-diagonalize $\b{H}^\text{4c}_\text{SF}$ and retain its large component block.
The same FW transformation is subsequently used on $\b{H}^\text{4c}_\text{SD}$ and on the Coulomb and Gaunt terms to obtain the complete X2C-1 Hamiltonian.

Without going into the detailed derivation which can be found for example in Ref.~\citenum{Liu2009}, it can be shown that the following FW transformation matrix block-diagonalizes the spin-free modified Dirac equation
\begin{align}
    \b{U} = \begin{pmatrix} \b{1} &\bar{\b{X}}\\ \b{X}& \b{1}\end{pmatrix}
    \begin{pmatrix} \b{R} &\b{0}\\ \b{0}& \bar{\b{R}}\end{pmatrix}
\label{eq:U}
,\end{align}
where the first matrix does the decoupling and the second renormalizes the states to bring them to the Schr\"{o}dinger metric. Only $\b{X}$ and $\b{R}$ are required to determine the large component of the block-diagonalized modified Dirac Hamiltonian. 
The matrix $\b{X}$ is defined through the relation 
\begin{align}
\phi^\text{L}_+=\b{X}\psi^\text{L}_+ \label{eq:X}
\end{align}
between the large component $\psi_+^\text{L}$ and the pseudo large component $\phi_+^\text{L}$ of the positive energy bispinors
\begin{align}
\Psi_+ = \begin{pmatrix}
 \psi_+^\text{L}\\
 \phi_+^\text{L}
\end{pmatrix}
\end{align}
and the matrix $\b{R}$ is defined as
\begin{align}
&\b{R}       = (\b{S}^{-1}\bar{\b{S}})^{-1/2}                    \label{eq:R} \\
&\bar{\b{S}} = \b{S} + \frac{\alpha^2}{2}\b{X}^\dag \b{T} \b{X}.
\end{align}
Using this unitary transformation, one can show that the spin-free two-component X2C Hamiltonian $\b{H}^\text{X2C}_\text{SF}$ is given as
\begin{align}
    \b{H}^\text{X2C}_\text{SF} = \b{R}^\dag (\b{V}_\text{ne} + \b{T}\b{X} + \b{X}^\dag \b{T} + \b{X}^\dag (\b{W}_\text{SF} - \b{T}) \b{X})\b{R} \label{eq:X2Csf}
.\end{align}
The same FW transformation is subsequently applied to the spin-dependent four-component Hamiltonian $\b{H}^\text{4c}_\text{SD}$ to obtain the one-body spin-orbit coupling operator
\begin{align}
    \b{H}^\text{X2C-1}_\text{SO}{}^{(1)} = -i\frac{\alpha^2}{4} \sum_{Ai} (\b{X}\b{R})^\dag\left(\b{p}_i \frac{Z_{ai}}{r_{Ai}} \times\b{p}_i\right)\cdot \B{\sigma}_i (\b{X}\b{R}) \label{eq:X2Cso1}
,\end{align}
which involves integrals
\begin{align}
    \label{eq:Sx2c}
    {}^\text{X2C}\!S_{pq}^\lambda=\sum_{rs}(\b{X}\b{R})^\dag_{pr}
    {}^\text{BP}\!S_{pq}^\lambda
   %\left(\sum_{iA}\sum_{\mu\nu} \epsilon_{\mu\nu}^\lambda \langle r_\mu|\frac{Z_{A}}{r_{iA}} | s_\nu\rangle \right)
    (\b{X}\b{R})_{sq}
,\end{align}
that are equal to the integrals found in the Breit-Pauli case transformed by the regularizing matrices $\b{X}$ and $\b{R}$.

\subsubsection{One-body part: X2C-N scheme}
An alternative scheme is to solve the complete modified Dirac equation found in Eq.~(\ref{eq:modDirac}), including both the spin-free and spin-dependent parts, to calculate the corresponding complex-valued FW transformation. The expressions for calculating the matrices $\b{X}$ and $\b{R}$ are formally equivalent to Eqs.~(\ref{eq:X}) and (\ref{eq:R}) respectively, with the difference that these are now $2n\times 2n$ dimensional matrices. The $\b{X}$ and $\b{R}$ matrices can be used in Eqs.~(\ref{eq:X2Csf}) and (\ref{eq:X2Cso1}) to obtain the complete two-component one-body hamiltonian containing both the spin-free and spin-dependent terms
\begin{align}
    \b{H}^{\text{X2C-N}} = \b{H}^{\text{X2C-N}}_{\text{SF}} + \b{H}^{\text{X2C-N}}_\text{SO}{}^{(1)}.\label{eq:X2Cson}
\end{align}
It is worth pointing out that the $\b{H}^{\text{X2C-N}}_{\text{SF}}$ contains contributions from high powers of $\b{W}_\text{SD}$ and is not the same as the genuine spin-free Hamiltonian $\b{H}^{\text{X2C}}_{\text{SF}}$ of Eq.~(\ref{eq:X2Csf}). However, as expression in Eq.~(\ref{eq:x2ceq}) indicates, we nevertheless use $\b{H}^{\text{X2C}}_{\text{SF}}$ in the X2C-N scheme. 

%Note that these spin-orbit coupling in X2C-N theory includes ccorrection up to infinite order, however somewhat unexpectedly the result obtained from them are  not always the most accurate because the four different components of the final Hamiltonian (including spin-free and spin-dependent one- and two-body terms) are often calculated at different orders and some inconsistencies are built into the approach\cite{?inconst}.

\subsubsection{Two-body part}
The two-body operator and its mean field approximation for the X2C theory are derived in great detail in Ref.~\citenum{Li2012} and ~\citenum{Li2014} and we merely give a brief outline of the derivation.

The X2C two-body operators are obtained by performing the same two sequential transformations (the restricted kinetic balance transformation and the spin-free unitary transformation) described in the ``One-body part: X2C-1 scheme'' section.
As a result, one obtains several terms, the first of which is the modified spin-free Coulomb term. This term is customarily replaced by the bare Coulomb term and the difference is ignored. Out of the remaining spin-dependent terms only the spin-same-orbital and spin-other-orbital are retained in our work. These are similar to their Breit-Pauli counterparts multiplied by the regularizing matrices $\b{X}$ and $\b{R}$.

The mean field approximation to the two-body part is derived by performing the X2C transformation on the spin-orbit contribution of the two-electron integrals in the modified Dirac-Fock-Coulomb-Breit operator\cite{Li2014}.
The resulting X2C mean field operator is
\begin{align}
    \b{H}^\text{X2C}_\text{MF}{}^{(1)}=i\frac{\alpha^2}{4} \sum_{pq} {}^\text{X2C}\!\bar{S}^\lambda_{pq} \hat{\sigma}_{ps}^\lambda
    \label{eq:X2Csomf}
,\end{align}
where the one-body integrals are
\begin{align}
    {}^\text{X2C}\!\bar{S}^\lambda_{pq} &= R_{pr} \left( \b{g}^{\text{LL},\lambda} + \b{g}^{\text{LS},\lambda} + \b{g}^{\text{SL},\lambda} + \b{g}^{\text{SS},\lambda} \right)_{rs}R_{sq}
    \label{eq:Sx2cmf}
.\end{align}
The matrices $\b{g}$ are
\begin{align}
    g^{\text{LL},\lambda}_{mn} =& -\sum_{pq}        2K^\lambda_{pmqn}                                             \gamma^\text{SS}_{pq}             \nonumber\\
    g^{\text{LS},\lambda}_{mn} =& -\sum_{pqr}  \left(K^\lambda_{mpqr} + K^\lambda_{pmqr}                   \right)\gamma^\text{LS}_{pq} X_{rn}      =g^{\text{SL},\lambda}_{nm}\label{eq:gx2c}\\
    g^{\text{SS},\lambda}_{mn} =& -\sum_{pqrs}2\left(K^\lambda_{rspq} + K^\lambda_{rsqp} - K^\lambda_{rpsq}\right)\gamma^\text{LL}_{pq} X^\dag_{mp}X_{qn}\nonumber
,\end{align}
the matrices $\b{X}$ and $\b{R}$ are formally defined in Eq.~(\ref{eq:X}) and (\ref{eq:R})
and the matrices $\B{\gamma}$ are
\begin{align}
    &\B{\gamma}^\text{LL} = \b{R}^\dag \B{\gamma}^\text{SA} \b{R}\\
    &\B{\gamma}^\text{LS} = \B{\gamma}^\text{LL} \b{X} =\B{\gamma}^\text{SL\dag}  \label{eq:gamma}\\
    &\B{\gamma}^\text{SS} = \b{X}^\dag \B{\gamma} \b{X}.
\end{align}
where $\B{\gamma}^\text{SA}=\tfrac{1}{2}(\B{\gamma}^\alpha+\B{\gamma}^\beta)$ is the spin-averaged density matrix. 
Finally, the two-electron integrals appearing in Eq.~(\ref{eq:gx2c}) are defined as
\begin{align}
K^\lambda_{pqrs} = \sum_{\mu\nu}\epsilon^\lambda_{\mu\nu}\langle p_\mu r_\nu |qs \rangle
\label{eq:Kx2c}
.\end{align}
The mean field X2C Hamiltonian exactly reduces to the mean field BP Hamiltonian when $\b{R} = \b{X} = \b{I}$. 

\section{Heat-bath Configuration Interaction}
Heat-bath Configuration Interaction (HCI) is a recently-developed\cite{HolTubUmr-JCTC-16} variant of the general class of methods that perform a selected configuration interaction calculation followed by perturbation theory (SCI-PT)\cite{Ivanic2001,Huron1973,Buenker1974,Evangelisti1983,Harrison1991,Steiner1994,Wenzel1996,Neese2003,Abrams2005,Bytautas2009,Knowles2015,Schriber2016,Liu2016,Caffarel2016,yann2017}. HCI, similar to other SCI-PT methods, consists of a variational step and a perturbative step. In the variational step, a set of important determinants is iteratively identified and the Hamiltonian is diagonalized in the space of these determinants. In the perturbative step, the energy obtained during the variational step is corrected using Epstein-Nesbet perturbation theory.
For details of the algorithm we refer the reader to our recent publications\cite{HolTubUmr-JCTC-16,ShaHolUmr-JCTC-17,holm17,hcicasscf}, here we briefly describe the key aspects of the method relative to other SCI-PT approaches.

\subsection{The heat-bath criterion}
At each iteration of the variational stage, the multireference ground state wavefunction
\begin{align}
    |\Psi\rangle = \sum_{D_i\in\mathcal{V}} c_i |D_i\rangle
\end{align}
with energy $E_0$ is calculated by diagonalizing the Hamiltonian in the current space $\mathcal{V}$ of important determinants $|D_i\rangle$ to obtain their coefficient $c_i$. The space $\mathcal{V}$ is augmented with a set of new determinants $|D_a\rangle$ that satisfy the HCI criterion
\begin{align}
    \max_{D_i \in \mathcal{V}} \left| H_{ai} c_i\right| > \epsilon_1, \label{eq:hci}
\end{align}
where $H_{ai}=\langle D_a|\hat{H}|D_i\rangle$ and $\epsilon_1$ is a user-defined parameter. The HCI criterion is different from the one used in CIPSI\cite{Huron1973,Evangelisti1983},
which is based on the contribution of a determinant $D_a$ to the perturbative correction to the wavefunction
\begin{align}
  \left|\frac{\sum_{\ket{D_i}\in \mathcal{V}}H_{ai}c_i }{E_0-E_a}\right| > \epsilon_1.
\end{align}
Although the HCI criterion is in principle suboptimal at picking out the most important determinants, in practice the difference is minimal.
Moreover, this slight difference is more than made up for by the significant advantage of the HCI criterion, which is that one generates \textit{only} the determinants $|D_a\rangle$ that satisfy the criterion in Eq.~(\ref{eq:hci}), and no resources are spent on generating determinants that would have been discarded.
The same criterion can also be used to speed up perturbation theory step.

\subsection{Stochastic perturbation theory}

The Epstein-Nesbet perturbative correction is evaluated as
\begin{align}
	E_2(\epsilon_2)=
    \sum_{\ket{D_a}}\frac{1}{E_0-E_a}
    \left(\sum_{\ket{D_i}\in\mathcal{V}}^{(\epsilon_2)} H_{ai}c_i\right)^2, \label{eq:pt_hci}
\end{align}
where the symbol $\sum^{(\epsilon_2)}$ designates a ``screened sum'' in which terms smaller in magnitude than a user-defined parameter $\epsilon_2$ are discarded and where the first sum is over the determinants $|D_a\rangle$ that meet this criterion. The exact perturbation correction is recovered in the limit of $\epsilon_2 \rightarrow 0$, and an $\epsilon_2$ much smaller than $\epsilon_1$ is needed in order to obtain a good approximation to the perturbative correction.	

Although the screened sum allows one to discard a large fraction of determinants, in most cases this is not sufficient to eliminate the memory bottleneck of having to store all the determinants $|D_a\rangle$ and their perturbative contributions in memory. To overcome this memory bottleneck, we use semistochastic perturbation theory\cite{ShaHolUmr-JCTC-17} to estimate the perturbative correction. In the semistochastic perturbation theory, an initial deterministic calculation
is performed with a relatively loose parameter $\epsilon_2^\text{d}$ to obtain an approximate perturbative correction $E_2^\text{D}(\epsilon_2^\text{d})$. The error in this calculated energy is then corrected stochastically by performing several iterations in which a stochastic perturbative correction is evaluated using the loose parameter $\epsilon_2^\text{d}$ and a much tighter parameter $\epsilon_2$. The near-exact perturbative correction with the tight parameter $\epsilon_2$ can then be calculated as 
\begin{align}
    E_2(\epsilon_2) =  E_2^\text{D}(\epsilon_2^\text{d}) + \bigg[ E_2^\text{S}(\epsilon_2) -  E_2^\text{S}(\epsilon_2^\text{d})\bigg] \label{eq:semistoc}
.\end{align}
The key to the success of the semistochastic perturbation theory is that both the loose and tight stochastic perturbative corrections, $E_2^\text{S}(\epsilon_2^\text{d})$ and $E_2^\text{S}(\epsilon_2)$, are calculated with the exact same set of sampled determinants. This correlated sampling significantly reduces the variance in the estimated perturbative corrections, thereby reducing the stochastic error.

\subsection{Spin-Orbit coupling and excited states}
With the introduction of the spin-orbit coupling, the Hamiltonian does not commute with the $\hat{S}_z$ operator and consequently $\langle \hat{S}_z\rangle$ is no longer a good quantum number. The SOC terms introduce non-zero matrix elements between determinants that contain different numbers of $\alpha$ and $\beta$ electrons, and thus break the degeneracy between the $2S+1$ states of a spin multiplet. The energy splitting can be measured experimentally to determine the zero-field splitting (ZFS), and
these energy differences are usually a small fraction of the absolute energies of the molecule. Thus two complications need to be addressed with the HCI algorithm: forming the wavefunction consisting of determinants with different numbers of $\alpha$ and $\beta$ electrons, and the special care needed to calculate the small energy differences with sufficient accuracy.

To address the first complication, let us recall that the variational step of the HCI algorithm includes three operations, identifying the important determinants to add to the space $\mathcal{V}$, finding the non-zero Hamiltonian matrix elements between all the determinants of $\mathcal{V}$ and diagonalizing the Hamiltonian. Each of these three operations are modified due to the addition of the SOC terms.
In identifying the important determinants, one now \emph{also} has to include those that can be generated with an $\alpha\rightarrow\beta$ or a $\beta\rightarrow\alpha$ excitation. On the other hand, the advantageous search protocol for doubly excited determinants that meet the HCI criterion does not change because the SOC terms only contain one-body operators. The Hamiltonian is stored in the sparse storage format and most of the terms in it are constructed using the
same algorithm as before\cite{ShaHolUmr-JCTC-17}, which avoids having to perform the expensive double loop over all determinants in the space $\mathcal{V}$. Additionally, matrix elements that break $\hat{S}_z$ symmetry are identified for each determinant by looping over all possible $\alpha$-occupied $\beta$-unoccupied and $\beta$-occupied $\alpha$-unoccupied orbital pairs to generate new determinants and doing a binary search over $\mathcal{V}$ to check if the determinant is present. Finally, the generalization of the Davidson algorithm to diagonalize the complex-valued Hamiltonian poses no serious problem.

Calculating the small energy splitting accurately requires that the energy of each state is calculated with a similar accuracy. The standard procedure for doing so is the state-average algorithm and was recently introduced in the context of HCI by modifying the HCI criterion of Eq.~(\ref{eq:hci})\cite{holm17}. In this work we introduce yet another different criterion that is more suitable for targeting degenerate states. The updated HCI criterion is 
\begin{align}
    \begin{array}{ll}
        &\max_{D_i \in \mathcal{V}} \left| H_{ai} \bar{c}_{i}\right| > \epsilon_1\\[.7em]
        &\bar{c}_{i} = \sqrt{\sum_{n} \left|c_{i}^n\right|^2}
    \end{array}
\label{eq:hciecited}
\end{align}
where $\bar{c}_{i}$ is the square root of the sum of the squares of the coefficients $c_{i}^n$ of determinant $|D_i\rangle$ in the $n^{th}$ state. This ensures that unitary transformations between degenerate states do not change the determinants that are added to $\mathcal{V}$.

\section{Computational details}\label{sec:comp}

Several different combinations of one-body scalar relativistic, two-body scalar relativistic, one-body SOC and two-body mean field SOC terms can be used. Here we perform calculations with 5 different Hamiltonians where the two-body scalar relativistic term is the Coulomb interaction while the other three terms are specified below: 
%The spin-orbit coupling calculations are done using combinations of the one- and two-body parts of the BP, X2C-1 and X2C-N Hamiltonian described above, yielding the following five distinct methods:

\begin{itemize}
    \item ``bp-bp''    uses the scalar relativistic, one-body and mean field two-body parts of the Breit-Pauli Hamiltonian.
                       This involves the integrals in Eqs.~(\ref{eq:Sbp}) and (\ref{eq:Sbpmf}).
    \item ``x2cn-bp''  uses the X2C-1 scalar relativistic Hamiltonian (Eq.~(\ref{eq:X2Csf})), the one-body part of the X2C Hamiltonian derived with the X2C-N scheme and the two-body part of the Breit-Pauli Hamiltonian.
                       This involves terms in Eqs.~(\ref{eq:X2Cson}) and (\ref{eq:Sbpmf}).
    \item ``x2cn-x2c'' uses the X2C-1 scalar relativistic Hamiltonian (Eq.~(\ref{eq:X2Csf})), the one-body part of the X2C Hamiltonian derived with the X2C-N scheme and the two-body part of the X2C Hamiltonian.
                       This involves the terms in Eqs.~(\ref{eq:X2Cson}) and (\ref{eq:Sx2cmf}).
    \item ``x2c1-bp''  uses the X2C-1 scalar relativistic Hamiltonian (Eq.~(\ref{eq:X2Csf})), the one-body part of the X2C Hamiltonian derived with the X2C-1 scheme and the two-body part of the Breit-Pauli Hamiltonian.
                       This involves the integrals in Eqs.~(\ref{eq:Sx2c}), (\ref{eq:Sbpmf}).
    \item ``x2c1-x2c'' uses the X2C-1 scalar relativistic Hamiltonian (Eq.~(\ref{eq:X2Csf})), the one- an two-body parts of the X2C Hamiltonian derived with the X2C-1 scheme.
                       This involves the integrals in Eqs.~(\ref{eq:Sx2c}), and (\ref{eq:Sx2cmf}).
\end{itemize}

In all calculations we begin by performing state-average HCISCF\cite{hcicasscf} (which is a CASSCF-like method where FCI is replaced by HCI) calculations targeting the relevant states with an equal weight. These HCISCF calculations are performed with a scalar relativistic Hamiltonian and SOC terms are not included at this stage. The optimized active space and orbitals are then used to perform a one-step SHCI calculation where both the scalar relativistic and SOC terms are included. Thus the spin-orbit
coupling and electron correlation are treated on an equal footing during the correlated calculation. In some calculations the initial HCISCF calculation and the subsequent one-step SHCI calculation are performed with a different active space. This is reasonable because the aim of the HCISCF calculations is to generate orbitals that are equally weighted towards all the targeted states; the relevant SOC calculation is the one-step SHCI calculation. 

All calculations were performed with a combination of the PySCF\cite{Sun2017} and Dice codes. The BP and X2C integrals and the routines for performing HCISCF calculations are implemented in PySCF and the SHCI calculations are performed using the Dice code. The SOC integrals are implemented using the automated code generator of the \texttt{libcint} library\cite{Sun2015LibcintAE}, which provides an efficient implementation of the standard integrals and their nuclear derivatives. 

In the following it is important to note that SHCI has stochastic error associated with it, which is sometime shown in parentheses, but is in any case much smaller than the targeted zero-field splitting.

\section{Results}

We present results of calculations done on the halogen group atoms (F, Cl, Br, I), the coinage metals group atoms (Cu, Au) as well as on the Neptunyl(VI) dioxide radical, NpO$_2^{2+}$, using the different Hamiltonians described in Section~\ref{sec:comp}.

\begin{table}
    \caption{\label{tab:halogen} Zero-field splitting results (cm$^{-1}$) for the atoms of the halogen group using SHCI with the ANO basis and using several active spaces and spin-orbit coupling methods.
    The stochastic errors on the zero-field splitting are systematically lower than 1 cm$^{-1}$ and the numbers in brackets are our estimates of the error in ZFS relative to the fully converged results calculated using the extrapolation scheme (see text).
    The results are compared to previously reported data from Ref.\citenum{B401472N}, \citenum{Cao2017} and \citenum{Li2014} and from experimental data from Ref.\citenum{halogenref}.
 }
 \begin{tabular}{lP{10mm}P{12mm}L{14mm}P{3mm}P{14mm}}
\hline\hline
Method    & \multicolumn{3}{c}{Active Spaces}   & & Ref.\\
\hline
\\
F         & (4o,7e) & (87o,7e)  &               & & \\
\cline{1-4}
bp-bp     & 405     & 399       &               & & \multicolumn{1}{|c}{403$^{a}$} \\
x2c1-bp   & 404     & 397       &               & & \multicolumn{1}{|c}{398$^{b}$} \\
x2cn-bp   & 404     & 398       &               & & \multicolumn{1}{|c}{421$^{c}$} \\
x2c1-x2c  & 405     & 399       &               & & \multicolumn{1}{|c}{404$^{d}$} \\
x2cn-x2c  & 405     & 399       &               & &  \\
\\
Cl        & (4o,7e) & (95o,7e)  & (100o,17e)    & &  \\
\cline{1-4}
bp-bp     & 834     & 805       & 873           & & \multicolumn{1}{|c}{823$^{a}$} \\
x2c1-bp   & 825     & 804       & 867           & & \multicolumn{1}{|c}{876$^{b}$} \\
x2cn-bp   & 822     & 802       & 858           & & \multicolumn{1}{|c}{907$^{c}$} \\
x2c1-x2c  & 827     & 806       & 866(1)        & & \multicolumn{1}{|c}{882$^{d}$} \\ 
x2cn-x2c  & 825     & 804       & 865           & &  \\
\\
Br        & (4o,7e) & (95o,7e)  & (100o,17e)    & & \\
\cline{1-4}
bp-bp     & 3680    & 3655      & 3797          & & \multicolumn{1}{|c}{3404$^{a}$} \\
x2c1-bp   & 3407    & 3382      & 3432          & & \multicolumn{1}{|c}{3649$^{b}$} \\
x2cn-bp   & 3373    & 3346      & 3396          & & \multicolumn{1}{|c}{3723$^{c}$} \\
x2c1-x2c  & 3428    & 3403      & 3454(93)      & & \multicolumn{1}{|c}{3685$^{d}$} \\
x2cn-x2c  & 3394    & 3366      & 3415          & &  \\
\\
I         & (4o,7e) & (116o,7e) & (121o,17e)    & & \\
\cline{1-4}
bp-bp     & 8816    & 9346      & 9752          & & \multicolumn{1}{|c}{6961$^{a}$} \\
x2c1-bp   & 6951    & 7199      & 7453          & & \multicolumn{1}{|c}{7755$^{b}$} \\
x2cn-bp   & 6816    & 7049      & 7246          & & \multicolumn{1}{|c}{7752$^{c}$} \\
x2c1-x2c  & 7021    & 7277      & 7487(62)      & & \multicolumn{1}{|c}{7603$^{d}$} \\
x2cn-x2c  & 6886    & 7127      & 7379          & &  \\
\hline\hline
\end{tabular}
\footnotesize
\raggedright
\\ $^a$ Ref.~\citenum{B401472N}
\\ $^b$ EOM-CCSD(SOC) from Ref.~\citenum{Cao2017}
\\ $^c$ X2Cmmf-FSCCSD from Ref.~\citenum{Li2014}
\\ $^d$ Exp. from Ref.~\citenum{halogenref}
\end{table}

\begin{figure}[htbp!]
	\centering
	\caption{\label{fig:halogen_cvg}
    Convergence of the total energies (Ha+2604) of the $^2\text{P}_{1/2}$ and $^2\text{P}_{3/2}$ states (blue and red dots in the upper part of the graph) and the zero-field splitting (cm$^{-1}$+3685, black dots in the lower part of the graph) with the SHCI perturbative correction.
    The dotted lines are linear fits through these points which are used to extrapolate the energies to vanishing perturbation energies.
    There is a small noise (less than 1\%) associated with the ZFS which can be traced to a combination of stochastic noise and uncertainty of the SHCI energies. 
   %The ZFS results are converged to within wavelength units for all calculations.
	}
	\includegraphics[width=\linewidth]{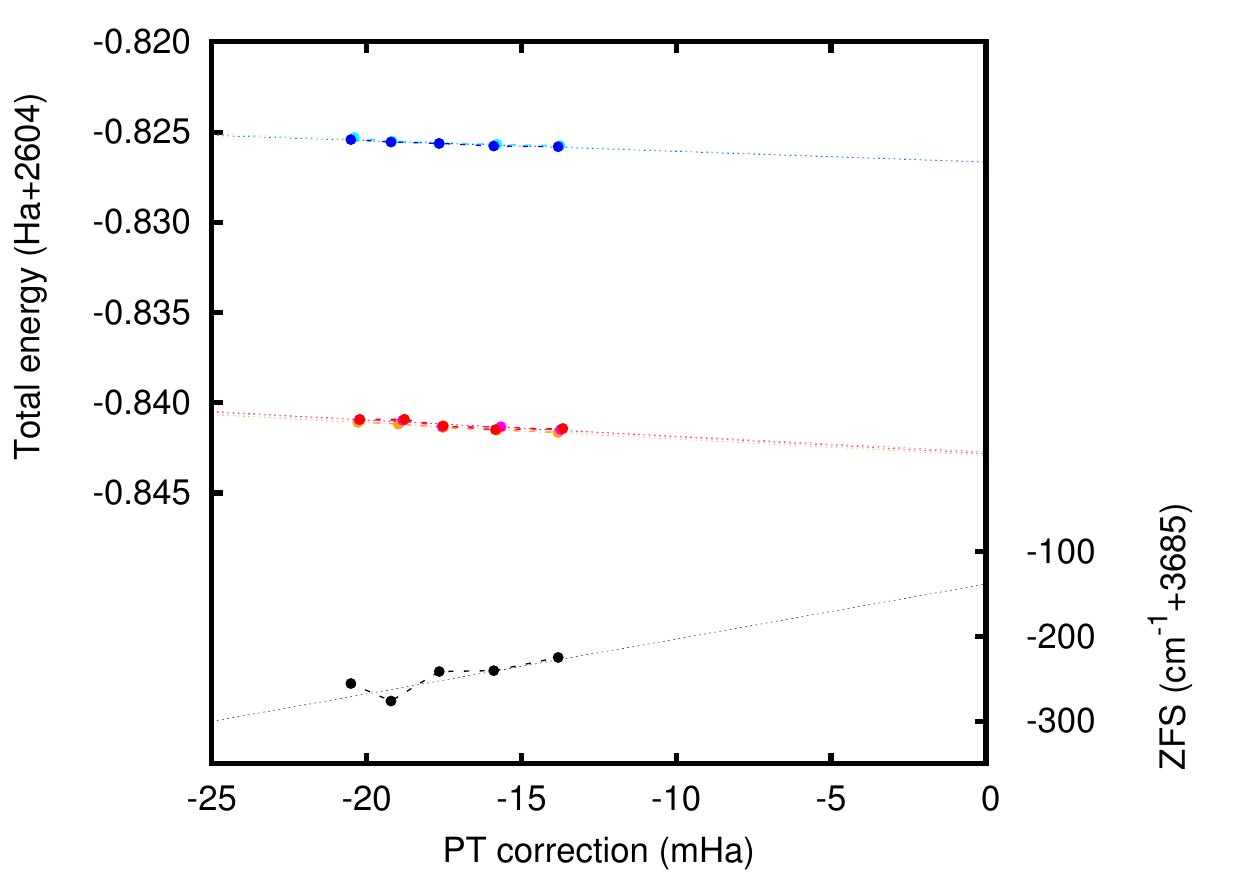}
\end{figure}

\subsection{Halogens}

We begin by performing a state-average HCISCF calculation where we target the three doubly-degenerate ground states corresponding to the singly occupied $p_x$, $p_y$ or $p_z$ orbitals using a valence active space of 4 electrons in 7 orbitals with the ANO basis set\cite{Roos2004}. We then perform a one-step SHCI calculation including SOC terms to calculate the six lowest energy states. We find a 4-fold degenerate
$^2\text{P}_{3/2}$ ground state and a 2-fold degenerate $^2\text{P}_{1/2}$ excited state, and their energy difference corresponds to the zero-field splitting measured experimentally and tabulated in the NIST reference table of Ref.~\citenum{halogenref} (displayed in the last column of Table \ref{tab:halogen}). 

%The (4o,7e) active space was used in all HCISCF calculations in which either the Schr\"{o}dinger equation or the scalar relativistic X2C Hamiltonian was used. Subsequently, five different one-step SHCI calculations are performed corresponding to the various combinations of scalar relativistic and SOC Hamiltonians.
It is important to note that the Hilbert space corresponding to the larger active spaces in Table~\ref{tab:halogen} can be enormous, for example the number of determinants that contribute to the ground and excited state of the Iodine atom with an active space of (121o,17e) is greater than 10$^{24}$ (this number is $^{242}C_{17}$, which assumes that no symmetry other than the particle number symmetry is used). With this kind of enormous active space, the individual energies of the targeted states
are not fully converged, however, the zero-field splitting is converged to better than 1\% of its FCI value except in the case of Br where it is accurate to 3\%. The estimated error in the energies is calculated using the extrapolation technique described in Ref. \citenum{holm17}, which uses the fact that near convergence the SHCI energies are nearly perfectly linear with respect to the SHCI perturbative correction. Such an extrapolation is shown in Figure~\ref{fig:halogen_cvg}, where this trend
seems to hold true in presence of the SOC terms as well. Such an extrapolation is used to estimate the error in the ZFS energies calculated using SHCI and is given in the bracket. %This extrapolation is only performed with the ``x2c1-x2c'' Hamiltonian. plotted against the perturbative correction and  ***CHECK THAT WITH NEW FIGURE (for now: problem in figures This can be seen from the result plotted in Figure~\ref{fig:halogen_cvg}, where the ZFS splitting converges to better than 5 cm$^{-1}$ for its final converged value of 3454 cm$^{-1}$ with the use of less than 2$\times 10^{6}$ determinants.*** This trend is found to be true in all calculations in this work. 

\begin{table*}
    \caption{\label{tab:coinage} Zero-field splitting results (eV) for the atoms of the coinage metal group using SHCI with the ANO basis and the ``x2c1-x2c'' spin-orbit coupling method.
    We report also some previous theoretical work, from Ref. \citenum{Malmqvist2002} and Ref.\citenum{Sayfutyarova2016}.     
 }
\begin{tabular}{lP{23mm}P{23mm}P{23mm}P{23mm}P{23mm}}
\hline\hline
States                                 & \multicolumn{4}{c}{Active Spaces}                                      & Ref.\cite{coinageref} \\
\hline
\\
Cu                                        & CASSCF-SO\cite{Malmqvist2002} & CASPT2-SO\cite{Malmqvist2002}    & DMRG-SISO\cite{Sayfutyarova2016}& SHCI  &      \\
                                        & (11o,11e) & (11o,11e) & (45o,19e)& (172o,19e) &      \\
\hline
$^2\text{D}_{5/2}$                         & 1.49      & 1.43      & 1.31& 1.39         & 1.39 \\
$^2\text{D}_{3/2}$                          & 1.75      & 1.69      & 1.57& 1.67        & 1.64 \\
$^2\text{D}_{5/2}- {}^2\text{D}_{3/2}$      & 0.26      & 0.26      & 0.26& 0.28        & 0.25 \\
\\
%Ag                                     & SHCI       &           &           &           &      \\
%                                       & (168o,29e) &           &           &           &      \\
%\hline
%$^2\text{D}_{5/2}$                     & 3.67       &           &           &           & 3.75 \\
%$^2\text{D}_{3/2}$                     & 4.25       &           &           &           & 4.30 \\
%$^2\text{D}_{5/2}-{}^2\text{D}_{3/2}$  & 0.58       &           &           &           & 0.55 \\
%\\
Au                                     & CASSCF-SO\cite{Malmqvist2002} & CASPT2-SO\cite{Malmqvist2002} & DMRG-SISO\cite{Sayfutyarova2016}& SHCI        &      \\%&  CASSCF-SO & DMRG-SISO  
                                        & (11o,11e) & (11o,11e) & (57o,43e) & (150o,25e)&      \\%&  (11o,11e) & (45o,43e)  
\hline                                                                                           %                          
$^2\text{D}_{5/2}$                           &  1.71     &  0.97     &  1.02&  1.15     & 1.14 \\%&  1.68      &  1.04      
$^2\text{D}_{3/2}$                           &  3.22     &  2.49     &  2.55&  2.64     & 2.66 \\%&  3.39      &  2.56      
$^2\text{D}_{5/2}- {}^2\text{D}_{3/2}$       &  1.51     &  1.52     &  1.53&  1.49     & 1.52 \\%&  1.71      &  1.52      
\hline\hline
\end{tabular}
\end{table*}

From Table~\ref{tab:halogen} it is possible to compare the performance of the various SOC Hamiltonians. We find that the ``x2c1-x2c'' scheme delivers the most accurate ZFS energies for most of the atoms. Interestingly, the simple Breit-Pauli results are extremely accurate for lower atomic weight species including Florine and Chlorine, however the results progressively start to deteriorate, significantly over-estimating the ZFS with the increase in the molecular
weight of the atom. This is to be expected because the Breit-Pauli SOC Hamiltonian is unbounded and thus over-estimates the energy splittings of a heavy atom in a variational calculation\cite{Dyall1997}.

The disagreement between the experimental and ``x2c1-x2c'' ZFS energies can be attributed to three causes, basis set incompleteness error, neglect of core-correlation and error in the relativistic Hamiltonian. For the Fluorine atom the core is fully correlated and it is surprising to note that the agreement with experiment becomes worse as we increase the active space to include the core electrons. This has to be attributed to basis-set incompleteness and deficiencies of the
relativistic treatment, which include neglect of various two-body scalar relativistic and spin-orbit coupling terms and the use of the SOMF approximation. At the moment it is not possible to disentangle the two because the current implementation of SHCI in Dice only works with the one-body SOC terms, however, in future publications we intend to address this point in more detail. Similar errors can be seen with other Halogens as well. In the cases of Bromine and Iodine, the improvement in agreement with the experiment when one increases the active space electrons from 7 to 17 suggests that including inner core electrons in the active space might further improve the agreement with experiments. 

\subsection{Coinage metals}

We have performed SHCI calculations on the more challenging coinage elements, Cu and Au. In all these calculations we again first perform a scalar relativistic state-average HCISCF calculation using Roos's ANO basis set\cite{BjornO.Roos2005}
with the 11 valence electrons in 11 orbitals
(including the valence $s$ and $d$ orbitals along with the virtual $d$ orbitals to account for the double-$d$ shell effect)
to simultaneously optimize with equal weight the 6 lowest lying states, 5 of which are degenerate, where the five 3$d$ and the 4$s$ orbital are singly occupied.
The obtained optimized orbitals are then used to perform a one-step SHCI calculation with the ``x2c1-x2c'' SOC Hamiltonian to obtain a 2-fold degenerate $^2\text{S}_{1/2}$ ground state, a 6-fold degenerate $^2\text{D}_{5/2}$ state and a 4-fold degenerate $^2\text{D}_{3/2}$ state. Two such calculations were performed, one in which the same (11o,11e) active space was used and another in which all the virtual orbitals were included in the active space in addition to semi-core electrons. We report the result of these calculations in Table~\ref{tab:coinage} which also contains the reference data from Ref.~\citenum{coinageref} and results of previous theoretical calculations.

\begin{figure}[htbp!]
	\centering
	\caption{\label{fig:coinage_fig}
    Errors (eV) in the individual $^2\text{D}_{5/2}$ and $^2\text{D}_{3/2}$ states of the coinage atoms (bars), as well as of the ZFS between the two (black line) of different methods previously reported and of SHCI calculations from this work.
    Although the ZFS is well reproduced by all methods, SHCI performs well in its description of the individual states.
	}
	\includegraphics[width=\linewidth]{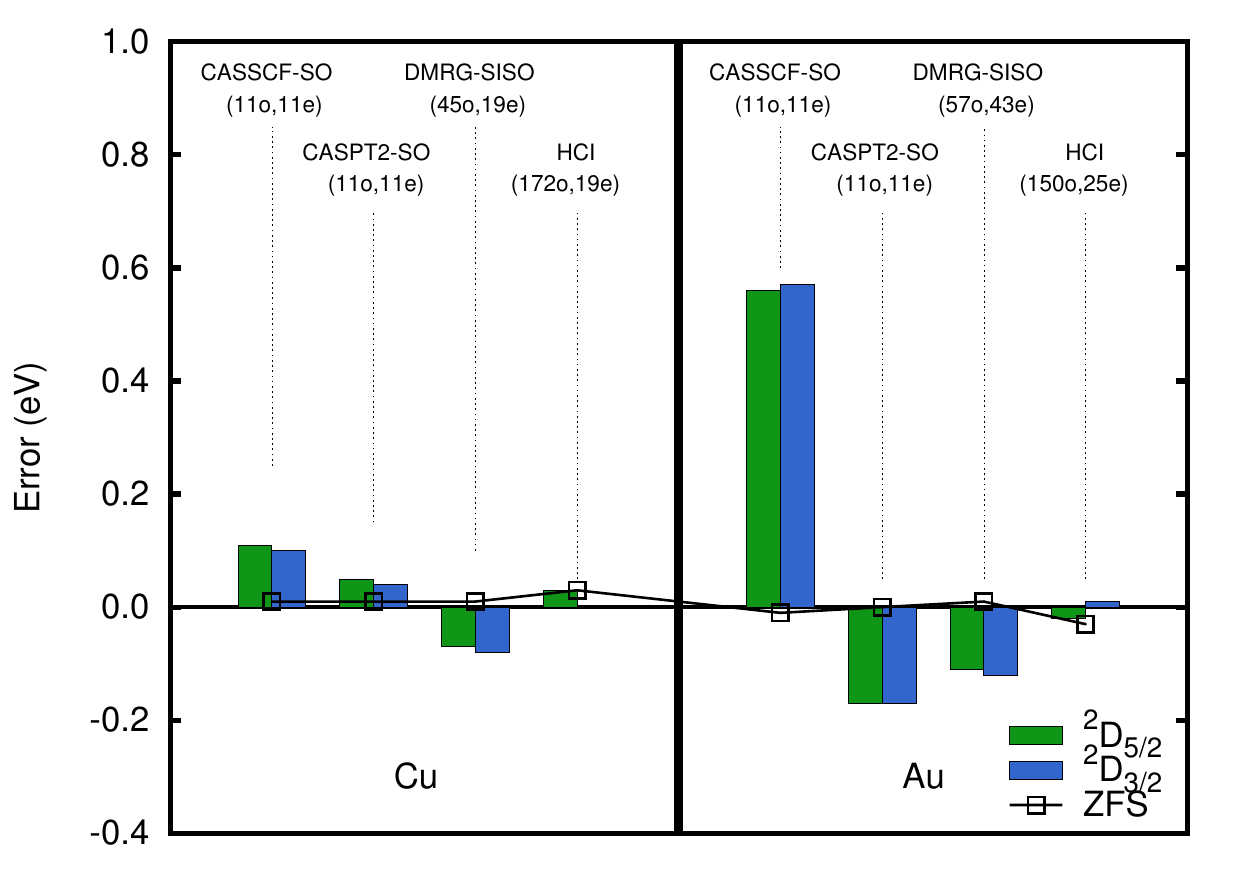}
\end{figure}

It is interesting to note (see Figure~\ref{fig:coinage_fig}) that although the ZFS splitting of the $^2\text{D}$ state into the $^2\text{D}_{5/2}$ and $^2\text{D}_{3/2}$ states, calculated by SHCI and other methods, are quite accurate, the excitation energies of these states relative to the $^2\text{S}$ ground state are very different. The agreement between the SHCI calculations and experiments is quite good with the maximum error being of just 0.03 eV, as opposed to other methods including the
large active space DMRG-SISO\cite{Sayfutyarova2016} calculations where the error can be as large as 0.12 eV. This difference in accuracy can be attributed to the use of the ANO-TZ basis set and the use of perturbative treatment of the SOC terms in the DMRG-SISO calculations. We note that the accuracy of CASPT2-SO\cite{Malmqvist2002} is significantly better than CASSCF-SO\cite{Malmqvist2002}, which points to the fact that dynamical and core correlations are important. %For all three atoms, some calculations have been reported that accurately predict the gap, such as CASPT2-SO and DMRG-SISO    . In those cases, though, this is a consequence of a rigid shift in the energies of the individual states obtained, with errors ranging from 0.12 to 0.57 eV in the case of the Gold atom. Such cancellations of errors are not needed for the SHCI calculations to reproduce the gaps with comparable accuracy to the DMRG-SISO calculation, but also obtain individual states that reproduce the reference data.

\begin{table}
    \caption{\label{tab:npo2pp}
    Relative energies of the electronic states of NpO$_2^{2+}$ calculated with ``x2c1-x2c'' SOC method for two different active spaces. We have also shown the results obtained previously using the atomic mean field spin-orbit Hamiltonian as implemented in Molcas and the SOC terms treated perturbatively.
 }
 \begin{tabular}{lR{20mm}R{20mm}R{15mm}}
\hline\hline
    States            & \multicolumn{2}{c}{Energies } & Ref.\cite{Gendron2014} \\
    \cline{2-3}
               &(4o,1e) & (143o,17e) & \\
\hline
    $^2\Phi  _{5/2}$      &0 &0  & 0    \\
    $^2\Delta_{3/2}$  & 4527& 3857     & 3011 \\
    $^2\Phi  _{7/2}$    &8568 & 8675  & 8092 \\
    $^2\Delta_{5/2}$   &10659& 10077   & 9192 \\
\hline\hline
\end{tabular}
\end{table}

\subsection{Neptunyl(VI)}

The ability to offer insight on magnetic and chemical properties of lanthanide and actinide containing complex found in single molecular magnets is important because experiments on those compounds prove difficult due to their high toxicity. Here, we perform relativistic calculations using the SHCI algorithm to calculate the ground and low lying excited states of NpO$_2^{2+}$, the Neptunyl(VI) dioxide radical\cite{Gendron2014a,Gendron2014,Knecht2016} using the ANO basis set truncated to the triple zeta
level. The molecule has $\text{C}_{\infty v}$ point group symmetry and the Np-O bond is of length 1.70 \AA, taken from the work of Gendron \textit{et al.}\cite{Gendron2014}. %Experimental data for these complexes do not exist and so we compare our results to those by Gendron. 
We begin by performing a state-average HCISCF calculation with a (4o,1e) active space in which the energies of the doubly degenerate $^2\Phi$ and $^2\Delta$ states are optimized using the X2C scalar relativistic Hamiltonian and we find an energy splitting equal to 2108 cm$^{-1}$. 
The energy difference between these two states is indicative of the strength of the crystal field splitting since in the absence of the axial oxygen atoms and relativistic effects, these states would be exactly degenerate.
Next we perform the one-step SHCI calculations with two different active spaces, (4o,1e) and (143o,17e), to obtain the 4 doubly degenerate states which are shown in Table~\ref{tab:npo2pp}. It is interesting to note that with the use of the same active space and geometry, the ZFS calculated here using the X2C Hamiltonian is significantly different from the one calculated by Gendron \textit{et al.}\cite{Gendron2014}, where the DKH Hamiltonian along with atomic mean field spin-orbit coupling as implemented in Molcas was used. The
ZFS splitting changes very little relative to this result even when slightly different geometry was used by Knecht \textit{et al.}\cite{Knecht2016}. The difference between our results and those of Gendron and Knecht can be attributed to the difference in the Hamiltonian used to treat the relativistic effect. For the Np atom, it will certainly be interesting to perform a full four-component calculation with the DCB Hamiltonian to calculate the ZFS and compare the accuracy of the various approximate
two-component Hamiltonians. Such an MRCI calculation was performed by Knecht, however only the calculated g-tensors was reported, and was in good agreement with the DKH results calculated there. It should be pointed out that even though the energy differences are quite different, we find that the ground state energy calculated with the (4o,1e) active space consists of 89\%  $^2\Phi$ state and 11\% $^2\Delta$ state, in agreement with the results of Knecht \textit{et al.}. This indicates
that the g-tensors calculated using SHCI would most likely have agreed well with those of DKH and four-component MRCI calculations performed by Knecht \textit{et al.}. 

%however SOC calculations, starting from a (4o,1e) HCISCF calculation to obtain orbitals offering a balanced description of the $^2\Phi$ ground-state and $^2\Delta$ first-excited state. ***DESCRIPTION OF STATES We then perform a one-step SHCI calculation including the SOC terms via the ``x2c1-x2c'' method described in the Computational Details, with the same active space, (4o,1e), as the HCISCF calculation and with a (143o,17e) active space. As seen in Table \ref{tab:npo2pp} and Figure \ref{tab:npo2pp}, the relative positions of the obtained split states are in good agreegment with reference data from Ref.\citenum{Gendron2014}.

\begin{figure}[htbp!]
	\centering
	\caption{\label{fig:npo22p}
    Energies of the electronic states of NpO$_2^{2+}$, relative to the SOC ground-state energy. On the left are spin-free and spin-orbit coupling states calculated using SHCI and the ``x2c1-x2c'' method, and an active space of (4o,1e) (black) and (143o,17e) (red), and on the right are reference data from Ref.\citenum{Gendron2014}.
}
	\includegraphics[width=\linewidth]{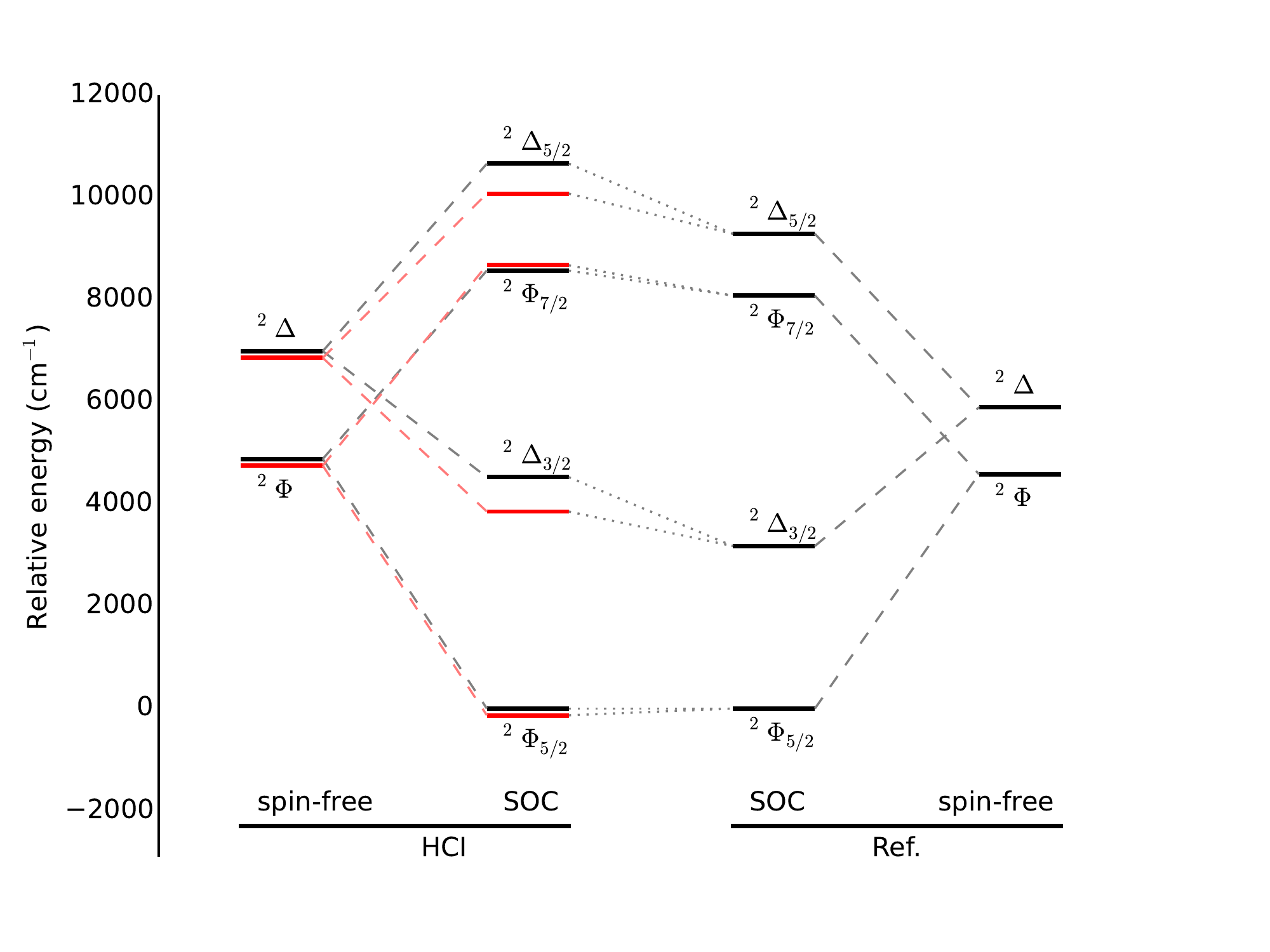}
\end{figure}

\section{Conclusion}
In this work we have extended the SHCI algorithm to treat the relativistic Hamiltonians containing spin-orbit coupling terms, by allowing one-body integrals and configuration interaction coefficients to be complex-valued numbers. The agreement with experimental results for several atomic species has been shown to be quite good and superior to previously calculated results. This is because not only are we treating large active spaces but we are also treating the electron correlation and
relativistic effects on an equal footing. The main shortcoming of the current methodology is that we cannot effectively include core correlation and treat larger systems because we do not have the ability to add dynamical correlation. Work in this direction is under way and we are looking for ways to combine SHCI with the recently-developed multireference linearized coupled cluster theory\cite{sharma2015multireference,Sharma2016b,Sharma2016}, which is formulated as a perturbation theory and uses Fink's partitioning\cite{Fink2006, Fink2009} of the Hamiltonian. We are also looking to extend the methodology so that the relativistic effects can already be included at the self-consistent field cycle to obtain complex-valued spinors. This will reduce the burden on the correlated calculations because a large part of the relativistic effects will be treated at the SCF level.
 
%We present in this work a new implementation in Dice where spin-orbit coupling and electron correlation are treated on an equal footing in a variational calculation. The zero-field splitting calculations we report are easily converged to within a few wavelength numbers using active spaces easily attainable for the HCISCF algorithm. This work introduces the development of spin-orbit coupling calculation as a new feature of HCISCF. Upcoming works will focus on the addition of dynamical correlation to this type of calculations using our recently-developed multireference linearized coupled cluster theory (MRLCC), which is formulated as a perturbation theory and uses the Fink’s partitioning of the Hamiltonian.

%\vspace*{1em}

\begin{acknowledgements}
The research was supported through the startup package from the University of Colorado, Boulder. We would also like to thank several useful discussions and insightful comments from Zhendong Li, Qiming Sun and Juergen Gauss.
\end{acknowledgements}

%%\bibliographystyle{biochem}
%%\bibliographystyle{jabbrv_ieeetr}
%\bibliographystyle{abbrv}
%\bibliography{paper,hci}

\end{document}